\documentclass[12pt,a4paper]{elsart}

\usepackage{epsfig}
\usepackage{amsmath}
\usepackage{cite}

\usepackage{xspace}
\newcommand{\aff}[2]{Dipartimento di Fisica dell'Universit\`a #1 e Sezione INFN, #2, Italy.}
\newcommand{\affd}[1]{Dipartimento di Fisica dell'Universit\`a e Sezione INFN, #1, Italy.}
\newcommand{\ke}{\ensuremath{K^\pm\to e^\pm\bar{\nu_e}}\xspace}
\newcommand{\km}{\ensuremath{K^\pm\to \mu^\pm\bar{\nu_\mu}}\xspace}

\begin{document}
\begin{frontmatter}
\title{Preliminary measurement of $\Gamma(\ke)/\Gamma(\km)$ at KLOE}
\collab{The KLOE Collaboration}
\author[Na]{F.~Ambrosino},
\author[Frascati]{A.~Antonelli},
\author[Frascati]{M.~Antonelli},
\author[Frascati]{F.~Archilli},
\author[Roma3]{C.~Bacci},
\author[Karlsruhe]{P.~Beltrame},
\author[Frascati]{G.~Bencivenni},
\author[Frascati]{S.~Bertolucci},
\author[Roma1]{C.~Bini},
\author[Frascati]{C.~Bloise},
\author[Roma3]{S.~Bocchetta},
\author[Roma1]{V.~Bocci},
\author[Frascati]{F.~Bossi},
\author[Roma3]{P.~Branchini},
\author[Roma1]{R.~Caloi},
\author[Frascati]{P.~Campana},
\author[Frascati]{G.~Capon},
\author[Na]{T.~Capussela},
\author[Roma3]{F.~Ceradini},
\author[Frascati]{S.~Chi},
\author[Na]{G.~Chiefari},
\author[Frascati]{P.~Ciambrone},
\author[Frascati]{E.~De~Lucia},
\author[Roma1]{A.~De~Santis},
\author[Frascati]{P.~De~Simone},
\author[Roma1]{G.~De~Zorzi},
\author[Karlsruhe]{A.~Denig},
\author[Roma1]{A.~Di~Domenico},
\author[Na]{C.~Di~Donato},
\author[Pisa]{S.~Di~Falco},
\author[Roma3]{B.~Di~Micco},
\author[Na]{A.~Doria},
\author[Frascati]{M.~Dreucci\corauthref{cor1}},
\author[Frascati]{G.~Felici},
\author[Frascati]{A.~Ferrari},
\author[Frascati]{M.~L.~Ferrer},
\author[Frascati]{G.~Finocchiaro},
\author[Roma1]{S.~Fiore},
\author[Frascati]{C.~Forti},
\author[Roma1]{P.~Franzini},
\author[Frascati]{C.~Gatti},
\author[Roma1]{P.~Gauzzi},
\author[Frascati]{S.~Giovannella},
\author[Lecce]{E.~Gorini},
\author[Roma3]{E.~Graziani},
\author[Pisa]{M.~Incagli},
\author[Karlsruhe]{W.~Kluge},
\author[Moscow]{V.~Kulikov},
\author[Roma1]{F.~Lacava},
\author[Frascati]{G.~Lanfranchi},
\author[Frascati,StonyBrook]{J.~Lee-Franzini},
\author[Karlsruhe]{D.~Leone},
\author[Frascati]{M.~Martini},
\author[Na]{P.~Massarotti},
\author[Frascati]{W.~Mei},
\author[Na]{S.~Meola},
\author[Frascati]{S.~Miscetti},
\author[Frascati]{M.~Moulson},
\author[Frascati]{S.~M\"uller},
\author[Frascati]{F.~Murtas},
\author[Na]{M.~Napolitano},
\author[Roma3]{F.~Nguyen},
\author[Frascati]{M.~Palutan},
\author[Roma1]{E.~Pasqualucci},
\author[Roma3]{A.~Passeri},
\author[Frascati,Energ]{V.~Patera},
\author[Na]{F.~Perfetto},
\author[Lecce]{M.~Primavera},
\author[Frascati]{P.~Santangelo},
\author[Na]{G.~Saracino},
\author[Frascati]{B.~Sciascia},
\author[Frascati,Energ]{A.~Sciubba},
\author[Pisa]{F.~Scuri},
\author[Frascati]{I.~Sfiligoi},
\author[Frascati]{A.~Sibidanov},
\author[Frascati]{T.~Spadaro},
\author[Roma1]{M.~Testa},
\author[Roma3]{L.~Tortora},
\author[Roma1]{P.~Valente},
\author[Karlsruhe]{B.~Valeriani},
\author[Frascati]{G.~Venanzoni},
\author[Frascati]{R.Versaci},
\author[Frascati,Beijing]{G.~Xu}

\address[Virginia]{Physics Department, University of Virginia, Charlottesville, VA, USA.}
\address[Frascati]{Laboratori Nazionali di Frascati dell'INFN, Frascati, Italy.}
\address[Karlsruhe]{Institut f\"ur Experimentelle Kernphysik, Universit\"at Karlsruhe, Germany.}
\address[Lecce]{\affd{Lecce}}
\address[Na]{Dipartimento di Scienze Fisiche dell'Universit\`a ``Federico II'' e Sezione INFN, Napoli, Italy}
\address[Energ]{Dipartimento di Energetica dell'Universit\`a ``La Sapienza'', Roma, Italy.}
\address[Roma1]{\aff{``La Sapienza''}{Roma}}
\address[Roma2]{\aff{``Tor Vergata''}{Roma}}
\address[Roma3]{\aff{``Roma Tre''}{Roma}}
\address[Pisa]{\affd{Pisa}}
\address[StonyBrook]{Physics Department, State University of New York at Stony Brook, NY, USA.}
\address[Beijing]{Permanent address: Institute of High Energy Physics, CAS, Beijing, China.}
\address[Moscow]{Permanent address: Institute for Theoretical and Experimental Physics, Moscow, Russia.}
\begin{flushleft}
\corauth[cor1]{cor1}{\small $^1$ Corresponding author: Mario Antonelli
INFN - LNF, Casella postale 13, 00044 Frascati (Roma), 
Italy; tel. +39-06-94032728, e-mail mario.antonelli@lnf.infn.it}
\end{flushleft}
\begin{flushleft}
\corauth[cor2]{cor2}{\small $^2$ Corresponding author: Tommaso Spadaro
INFN - LNF, Casella postale 13, 00044 Frascati (Roma), 
Italy; tel. +39-06-94032698, e-mail tommaso.spadaro@lnf.infn.it}
\end{flushleft}

\begin{abstract}
A preliminary measurement of $R = \Gamma(\ke)/\Gamma(\km)$ at the KLOE
experiment is discussed. The result,
$R = (2.55\pm0.05\pm0.05)\times10^{-5},$
is based on 1.7 fb$^{-1}$ of luminosity integrated on the $\phi$-meson peak at the
Frascati $e^+e^-$ collider DA$\Phi$NE, corresponding to
$\sim$8000 observed \ke events. Perspectives on the methods planned to improve both the statistical
and the systematic errors are briefly outlined.
\end{abstract}
\end{frontmatter}

\section{Introduction}
A strong interest for a new measurement of the ratio $R_K=\Gamma(\ke)/\Gamma(\km)$ has
recently arisen, triggered by the work of Ref.~\citen{masiero}. The
SM prediction of $R_K$ benefits from cancellation of hadronic uncertainties to a large extent and 
therefore can be calculated with high precision. Including radiative corrections, the total uncertainty is 
less than 0.5 per mil~\cite{finkemeier}.
Since the electronic channel is helicity-suppressed by the $V-A$ structure of the charged weak current,
$R_K$ can receive contributions from physics beyond the SM, for example from multi-Higgs effects inducing an effective pseudoscalar interaction.
It has been shown in Ref.~\citen{masiero} that deviations from the 
SM of up to few percent on $R_K$ are quite possible in minimal supersymmetric extensions
of the SM and in particular should be dominated by lepton-flavor violating contributions with tauonic neutrinos emitted. 
Using the present KLOE dataset of $\sim$2.5 fb$^{-1}$ of luminosity integrated at
the $\phi$-meson peak, we show that an accuracy of about 1~\% in the measurement of $R_K$ might be reached. 

In order to compare with the SM prediction at this level of accuracy, one has to treat carefully
the effect of radiative corrections, which contribute several percent to the $K_{e2}$ width. 
In particular, the SM prediction of Ref.~\cite{finkemeier} is made considering all photons emitted by
the process of internal bremsstrahlung (IB) while ignoring any contribution from structure-dependent direct emission (DE). 
Of course both processes contribute, so in the analysis we will consider DE as a background
which can be distinguished from the IB width by means of a different photon energy spectrum.

\section{Experimental setup}
DA$\Phi$NE, the Frascati $\phi$ factory, is an $e^{+}e^{-}$ collider
working at $\sqrt{s}\sim m_{\phi} \sim 1.02$~GeV. $\phi$ mesons are produced,
essentially at rest, with a visible cross section of $\sim$~3.1~$\mu$b
and decay into $K^+K^-$ pairs with a BR of $\sim 49$\%.

Kaons get a momentum of $\sim$~100~MeV/$c$ which translates into a low speed, $\beta_{K} \sim$ 0.2.
$K^+$ and $K^-$ decay with a mean length of $\lambda_\pm\sim $~90~cm and can be distinguished by their decays in flight to one of the two-body final states 
$\mu\nu$ or $\pi\pi^0$. Observation of a $K^+$ in an event signals, or tags, the presence of a $K^-$
and vice versa; highly pure and nearly monochromatic $K^\pm$
beams can thus be obtained and exploited to achieve high precision in the measurement of absolute BR's.

The analysis of kaon decays is performed with the KLOE detector, consisting essentially of a drift chamber, DC, surrounded by an
electromagnetic calorimeter, EMC. A superconducting coil provides a 0.52~T magnetic field.
The DC~\cite{DCnim} is a cylinder of 4~m in diameter
and 3.3~m in length, which constitutes a fiducial volume 
for $K_L$ and $K^\pm$ decays extending up to $\sim0.4\lambda_{L}$ and $\sim1\lambda_\pm$, respectively.
The momentum resolution for tracks 
at large polar angle is $\sigma_{p}/p \leq 0.4$\%. 
The distribution of c.m.\ momenta reconstructed from identification of one-prong $K^\pm\to\mu\nu,\pi\pi^0$ decay vertices in the DC shows a peak
around the expected value with a resolution of 1--1.5~MeV, thus allowing clean $K^\mp$ tagging. 

The EMC is a lead/scintillating-fiber sampling calorimeter~\cite{EmCnim}
consisting of a barrel and two endcaps, covering 98\% of the solid angle.
The EMC is readout at both ends by photomultiplier tubes. 
The PM signals provide the energy deposit magnitude. Their timing provide
the arrival times of particles and the three-dimensional positions of the
energy deposits are determined from the signal at the two ends. 
The readout granularity is $\sim4.4\times4.4$~cm$^2$, with 2440 ``cells'' 
arranged in layers five-deep. Cells close in time and space are grouped into a ``calorimeter cluster.'' For each cluster, the energy
$E_\mathrm{cl}$ is the sum of the cell energies, and the time $t_\mathrm{cl}$ and position $\vec{r}_\mathrm{cl}$ are calculated as
energy-weighted averages over the fired cells.
The energy and time resolutions are $\sigma_{E}/E \sim 5.7\%/\sqrt{\rm{E(GeV)}}$ and 
$\sigma_{T} =$~54~ps$/\sqrt{\rm{E(GeV)}} \oplus 50$ ps, respectively.
The timing capabilities of the EMC are exploited to precisely reconstruct the position of decay vertices of $K^\pm$ to $\pi^0$'s from the
cluster times of the emitted photons, thus allowing precise measurements of the $K^\pm$ lifetime.

In early 2006, the KLOE experiment completed data taking, having collected
$\sim2.5$~fb$^{-1}$ of integrated luminosity at the $\phi$ peak,
corresponding to $\sim$3.6 billion $K^+ K^-$ pairs. The preliminary result 
presented here is based on the analysis of 1700~pb$^{-1}$.

A Monte Carlo (MC) data set was produced on a run-by-run basis, with luminosity scale factors equal to 1
for the main $K^\pm$ decay channels and 100 for decay channels with BR's less than 10$^{-4}$.

\section{Event selection}
Given the $K^\pm$ decay length of $\sim$90~cm, the selection of one-prong $K^{\pm}$ decays in the DC required to tag $K^{\mp}$ has an efficiency smaller
than 50\%. In order to keep the statistical uncertainty on the number of \ke counts below 1\%, we decided to perform a ``direct search'' for \ke and
\km decays, without tagging. Since we measure a ratio of BR's for two channels with similar topology and
kinematics, we expect to benefit from some cancellation of the uncertainties on tracking, vertexing, and kinematic identification efficiencies.
Small deviations in the efficiency due to the different masses of electrons and muons can be evaluated using MC. 

Selection starts requiring the presence of a one-prong decay vertex of a kaon track in a fiducial volume (FV) in the DC consisting of
a track with laboratory momentum between 70 and 130~MeV, which can be extrapolated backward to a region near the interaction point, and 
a secondary track of relatively high momentum (between 180 and 270~MeV).
The FV is defined as a cylinder parallel to the beam axis with length of 80~cm, and inner and outer radii of 40 and 150~cm, respectively. 
Quality cuts are applied using $\chi^2$-like variables for the tracking of kaon and secondary particle and for the vertex fit.
These requirements are referred to as the ``one-prong selection'' in the following.

\begin{figure}[ht]
 \center
    \epsfig{file=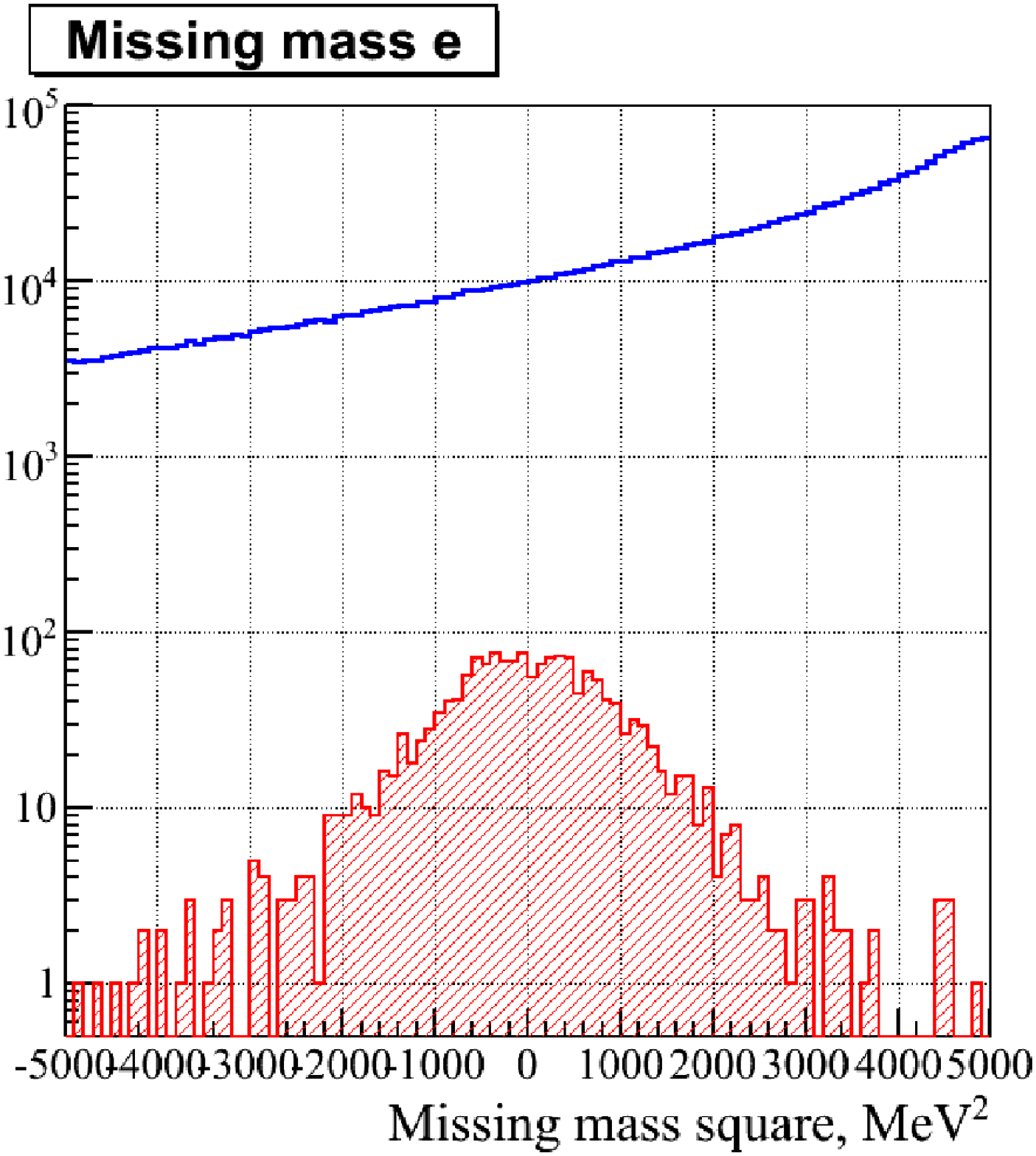, width=0.48\textwidth}
    \epsfig{file=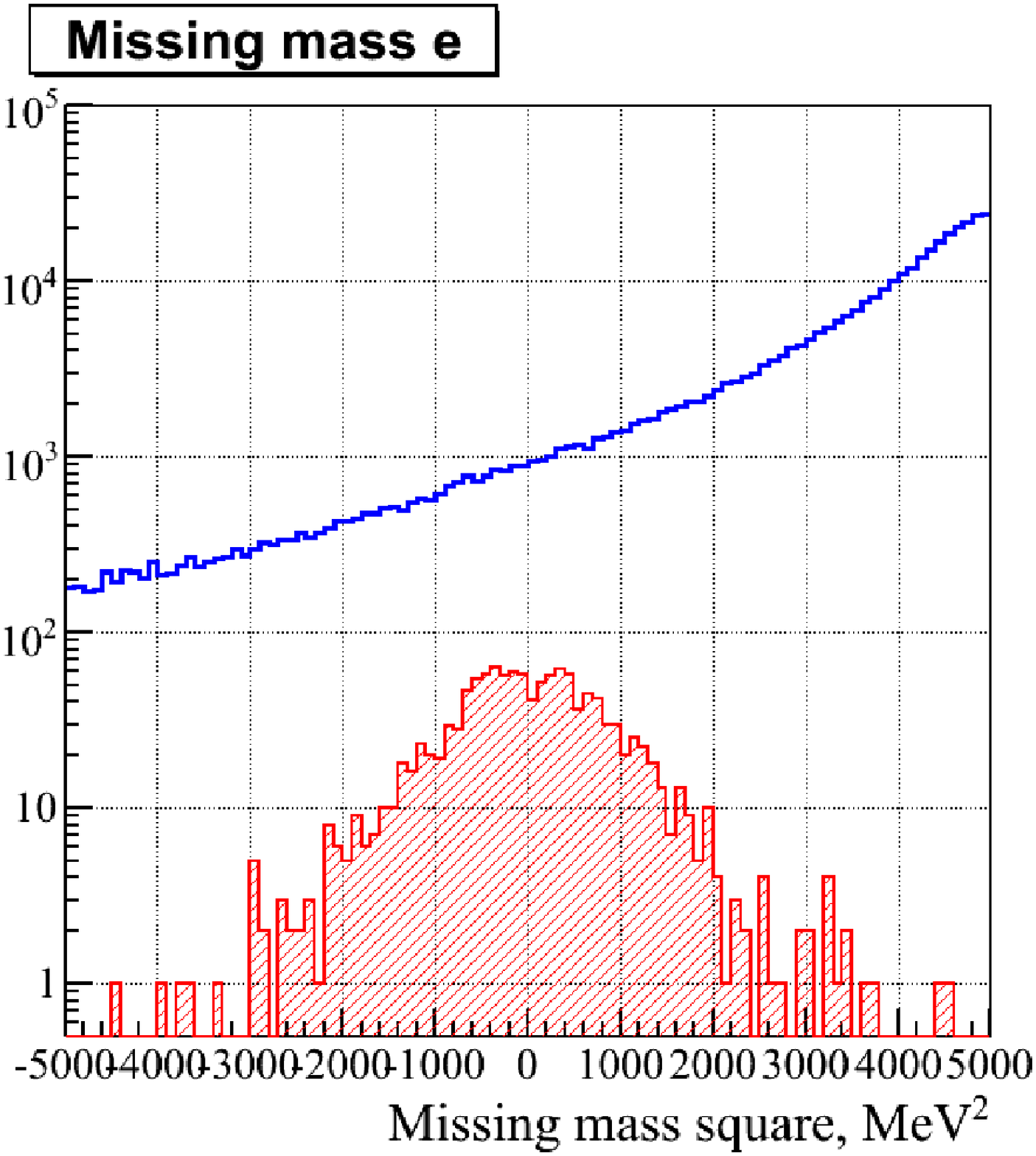, width=0.48\textwidth}
\caption{MC distribution of $M_\mathrm{lep}^2$ before (left) and after (right) quality cuts are applied. 
The shaded histograms correspond to \ke events and the open histograms to the background, which is dominated by \km events. In the MC,
the ratio $R_K$ is set to the SM value.}
\label{ke2:mlep}
\end{figure}

A powerful kinematic variable used to distinguish \ke and \km decays from the background is calculated from
the momenta of the kaon and the secondary particle measured in DC: assuming zero neutrino mass one can obtain the squared mass of the
secondary particle, or lepton mass ($M_\mathrm{lep}^2$). The distribution of $M_\mathrm{lep}^2$ is shown in Fig.~\ref{ke2:mlep} for MC events
before and after quality cuts are applied. While the one-prong selection is enough for clean identification of a \km sample,
further rejection is needed in order to identify \ke events: the background, which is dominated by badly reconstructed \km events, 
is reduced by a factor of $\sim$10 by
the quality cuts, but still remains $\sim$10 times more frequent than the signal in the region around the electron mass peak. The one-prong selection 
efficiency is $\sim28\%$ for both channels, and the ratio of efficiencies for \km and \ke is evaluated from MC to be:
$\epsilon^\mathrm{TRK}_{K\mu2}/\epsilon^\mathrm{TRK}_{Ke2}=0.974(1)$. A correction to this estimate 
accounting for possible differences in the tracking performance between data and MC is discussed in Sec.~\ref{ke2:result}.

\begin{figure}[ht]
 \center
    \epsfig{file=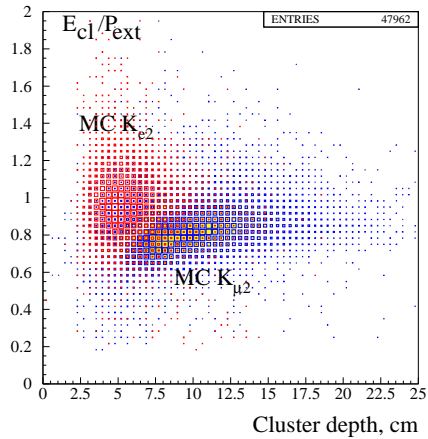, width=0.48\textwidth}
\caption{MC distribution of the ratio $E_\mathrm{cl}/P_\mathrm{ext}$ of cluster energy and track momentum 
as a function of the depth of the cluster along the direction of impact of the secondary particle on the EMC.}
\label{ke2:ell}
\end{figure}

Information from the EMC is used to improve background rejection. The secondary track is extrapolated to a position $\vec{r}_\mathrm{ext}$ on the 
EMC surface with momentum $\vec{p}_\mathrm{ext}$
and associated to nearest calorimeter cluster satisfying the impact-parameter cut $d_\perp<30$~cm, where
$d_\perp = |\vec{p}_\mathrm{ext}/|p_\mathrm{ext}|\times(\vec{r}_\mathrm{ext}-\vec{r}_\mathrm{cl})|$.
For electrons, the associated cluster is close to the EMC surface so that its position projected along the track 
$d_\parallel= |\vec{p}_\mathrm{ext}\cdot(\vec{r}_\mathrm{ext}-\vec{r}_\mathrm{cl})|$ is only a few cm. Moreover, for electrons the cluster energy 
$E_\mathrm{cl}$ is a measurement of the particle momentum $p_\mathrm{ext}$. Therefore the 
following condition is required 
in the plane $E_\mathrm{cl}/p_\mathrm{ext}$ vs $d_\parallel$ (see Fig.~\ref{ke2:ell}):
\begin{equation}
\left(\frac{d_\parallel[\mathrm{cm}]-2.6}{2.6}\right)^2+
\left(\frac{E_\mathrm{cl}/p_\mathrm{ext}-0.94}{0.2}\right)^2<2.5.
\end{equation}

Electron clusters can be further distinguished from $\mu$ (or $\pi$) clusters by exploiting
the granularity of the EMC: electrons shower and deposit their energy
mainly in the first plane of EMC, while muons behave like minimum ionizing
particles in the first plane while they deposit a sizable fraction of their kinetic energy from the third plane onward when they are slowed down to rest (Bragg
peak). The particle identification (PID) was therefore based on the asymmetry $A_f$ of energy deposits between the second and the first planes hit, 
on the spread $E_\mathrm{RMS}$ of energy deposits on each plane, on the position $r_\mathrm{max}$ of the plane with the maximum energy,
and on the asymmetry $A_l$ of energy deposits between the last and the next-to-last planes.
Muon clusters with the signature $A_f>0$, or $x_\mathrm{max}>12$~cm, or $A_l<-0.85$ are rejected.

\begin{figure}[ht]
 \center
    \epsfig{file=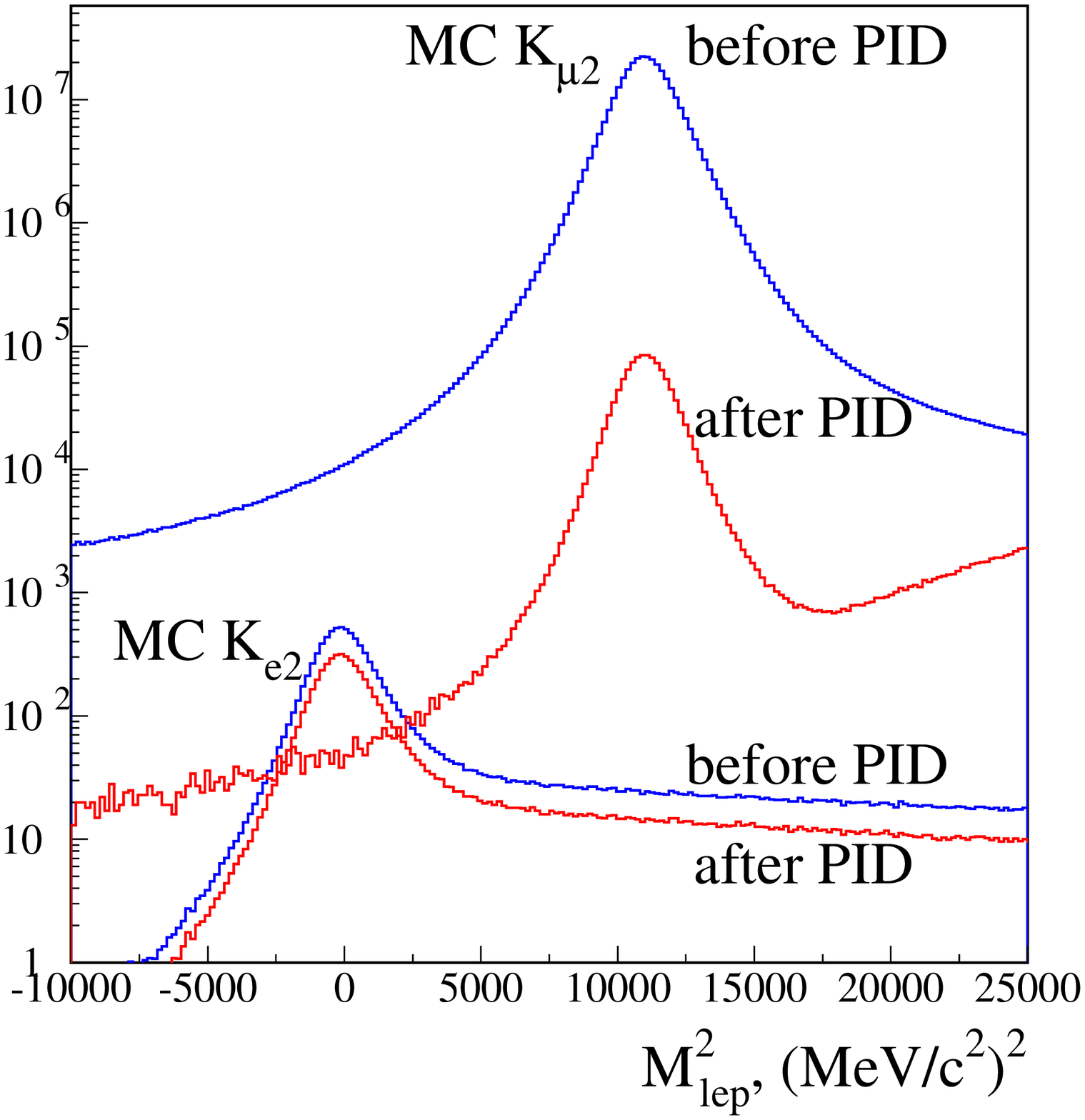, width=0.48\textwidth}
    \epsfig{file=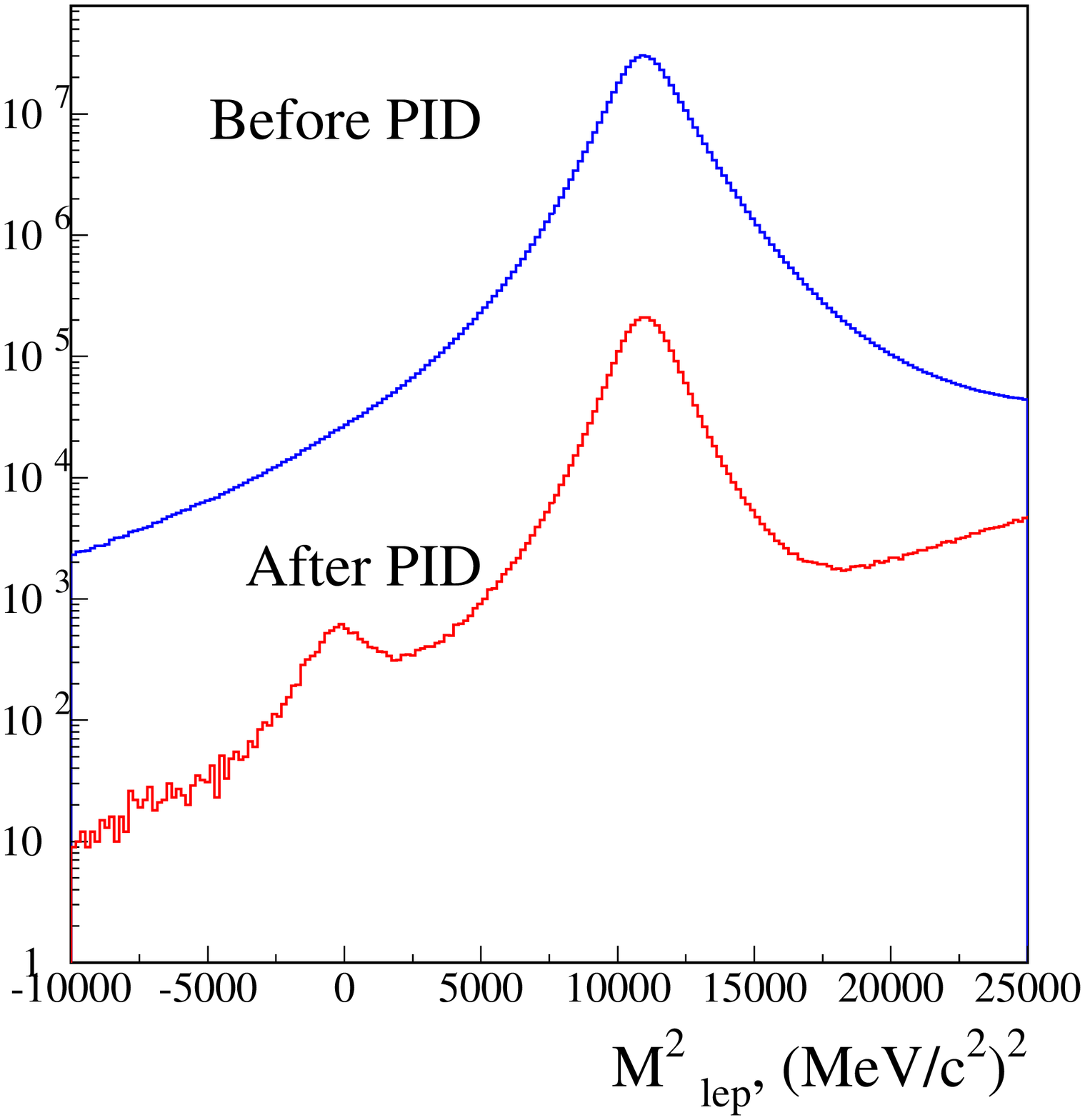, width=0.48\textwidth}
\caption{Left: MC distribution of $M_\mathrm{lep}^2$ for background (plots with larger population) and for signal (lower populations) before and 
after the PID is required. Right: data distribution of $M_\mathrm{lep}^2$ before and after the PID is required. The \ke signal is visible only
after PID.}
\label{ke2:PID}
\end{figure}

The PID technique described above selects \ke events with an efficiency $\epsilon^\mathrm{PID}_{Ke2}\sim64.7(6)\%$ and a rejection power for
background of about 300. These numbers have been evaluated from MC. The effect of the improvement in background rejection obtained with PID is
visible by comparing $M_\mathrm{lep}^2$ distributions before and after the PID is applied; see Fig.~\ref{ke2:PID}.

\section{Event counting}
\begin{figure}[ht]
 \center
    \epsfig{file=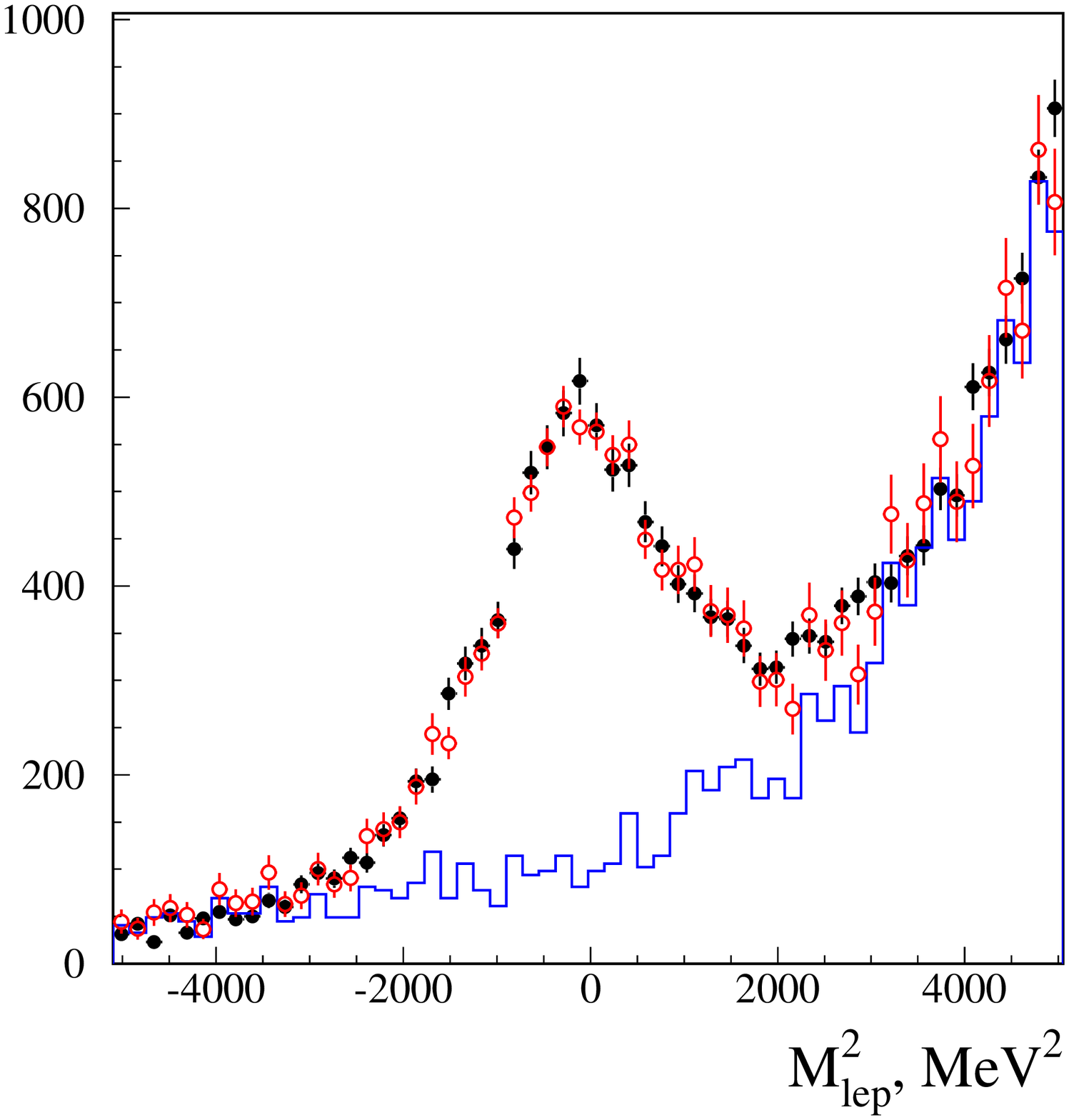, width=0.48\textwidth}
    \epsfig{file=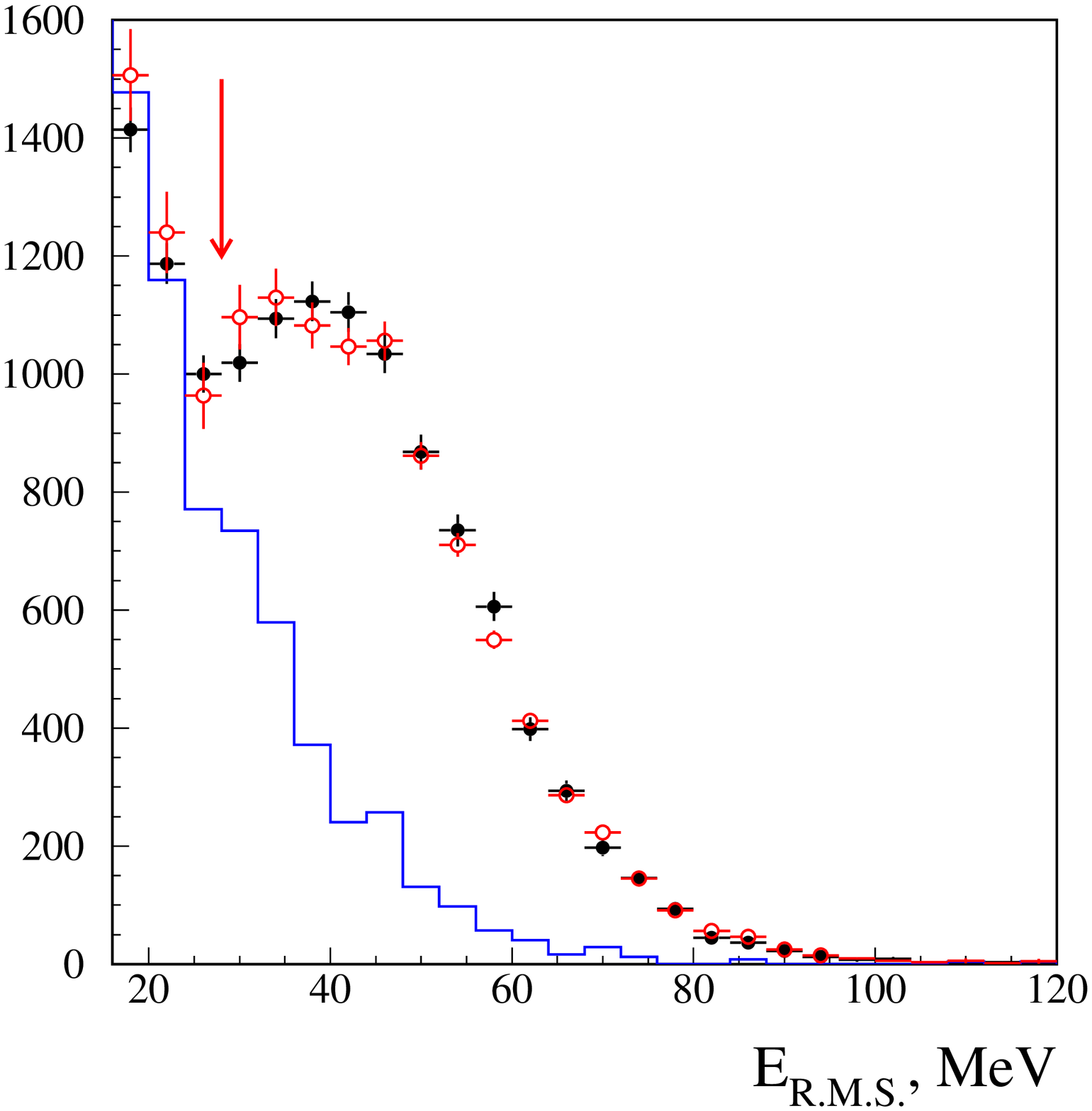, width=0.48\textwidth}
\caption{Distributions of the lepton mass squared $E^2_\mathrm{lep}$ of the secondary track (left panel) and of the spread $E_\mathrm{RMS}$ of 
the energy deposits among the planes of the connected cluster in the EMC (right panel). Filled dots represent the data, while open dots are the result from a
maximum-likelihood fit using signal and background (solid line) distributions as input from MC.}
\label{ke2:fit}
\end{figure}

A likelihood fit to the two-dimensional $E_\mathrm{RMS}$ vs 
$M_\mathrm{lep}^2$ distribution was performed to 
get the number of signal events. Distribution shapes
for signal and background were taken from MC; the normalizations for the two components are the only fit parameters. 
The number of signal events obtained from the fit is $N_{Ke2}=8090\pm156$. Projections of the fit results onto the two axes are compared to
real data in Fig.~\ref{ke2:fit}. 

\begin{figure}[ht]
 \center
    \epsfig{file=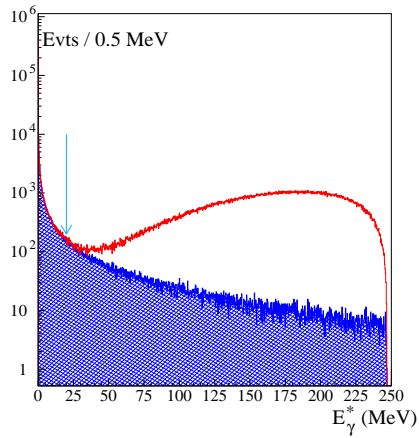, width=0.48\textwidth}
\caption{Distribution of the energy of the emitted photon in a $K\to e\nu\gamma$ decay, from the inner bremsstrahlung width (hatched histogram) or through
the total inner bremsstrahlung $+$ direct emission processes.}
\label{ke2:IBDE}
\end{figure}

The primary generators for \ke and \km decays include radiative corrections and allow for the emission of a single photon in the final state~\cite{radcor}. 
$\ke+\gamma$ events with photon energy in the kaon rest frame $E_{\gamma}<20$~MeV were
considered as signal: as shown in Fig.~\ref{ke2:IBDE}, the DE contribution is indeed negligible in this range.
The fraction of the IB component lying in the chosen energy range is determined from MC to be $\epsilon^\mathrm{IB}=0.9528(5)$.

While evaluating the shape for $K\to e\nu(\gamma)$, the present PDG value has been used for the
ratio of IB and DE contributions: $\mathrm{IB}/(\mathrm{IB}+\mathrm{DE})=0.50(4)$~\cite{PDG}.
The fit has been repeated with different values of this ratio, 
varied within its range of uncertainty. This procedure gave a $\sim$0.45\% error on the number of signal counts. 

The number of \km events in the same data set is extracted from
a similar fit to the $M_\mathrm{lep}^2$ distribution, where no PID cuts are applied in this case. The fraction of
background events under the muon peak is estimated from MC to be less than one per mil.
The number of \km events is 499\,251\,584$\pm$35403. 

\section{Evaluation of $R_K$}
\label{ke2:result}
The following formula has been used to evaluate the ratio $R_K$:
\begin{equation}
R_K = 
\frac{N_{Ke2}}{N_{K\mu 2}} 
\left[\frac{\epsilon^\mathrm{TRG}_{K\mu2}}{\epsilon^\mathrm{TRG}_{Ke2}}\right]
\left[C^\mathrm{TRK}
\frac{\epsilon^\mathrm{TRK}_{K\mu2}}{\epsilon^\mathrm{TRK}_{Ke2}}\right]
\left[\frac{1}{C^\mathrm{PID}\epsilon^\mathrm{PID}_{Ke2}}\right]
\frac{1}{\epsilon^\mathrm{IB}},\label{eq:ratio}
\end{equation} 
where $N_{Ke2}$ and  $N_{K\mu2}$ are the number of \ke and \km observed events;
$\epsilon^\mathrm{TRK}_{Ke2}$ and $\epsilon^\mathrm{TRK}_{K\mu 2}$ are the efficiencies for the one-prong selection
for \ke and \km decays, evaluated from MC; 
the correction $C^\mathrm{TRK}$ to their ratio accounts for possible differences between the data and the MC prediction.
The PID efficiency for \ke events $\epsilon^\mathrm{PID}_{Ke2}$ has been evaluated from MC, while a correction $C^\mathrm{PID}$ has been evaluated to 
account for possible discrepancies between data and MC in the description of PID variables.
Trigger efficiencies $\epsilon^\mathrm{TRG}$ were instead evaluated directly from data.
The estimates of the above efficiencies and corrections are briefly discussed in the following section.
Finally, $\epsilon^\mathrm{IB}$ is the fraction of the IB component accepted in the selection of \ke
events and has been evaluated from MC.

\subsection{Evaluation of the efficiency for one-prong selection}
The MC estimates for $\epsilon^\mathrm{TRK}$ have been checked using a control sample of \km events. This sample has been selected 
by requiring a tagging two-body kaon decay in the DC and by identifying a secondary muon cluster coming from a \km decay of the other
kaon. In this sample, the kaon decay vertex and the laboratory momentum 
of the emitted muon are evaluated using the information from the tag and from the selected cluster with resolutions of 5~cm and 13~MeV, respectively. 
We evaluated from this sample $\epsilon^\mathrm{TRK}$ for both data and
MC, as a function of the laboratory momentum of the secondary particle and the position of the one-prong vertex. 
Convoluting the data/MC ratio with the \ke and \km kinematics, we get a correction
$C_\mathrm{TRK}=0.994(9)$ for the ratio $\epsilon^\mathrm{TRK}_{K\mu2}/\epsilon^\mathrm{TRK}_{Ke2}$. Only 10~pb$^{-1}$ have been used to obtain
this estimate. The quoted statistical error will be reduced to a negligible level after processing of the entire statistics.

\subsection{Evaluation of the efficiency for PID}
In order to check the reliability of the MC for the efficiency estimate,
a control sample (CS) of $K_{Le3}$ decays has been selected and used to compare data with MC. 
Using 600 pb$^{-1}$ of integrated luminosity about 200k $K_{Le3}$ events with
a purity of 99.7\% have been selected. The ratio of data and MC PID efficiencies has been evaluated as function of
the electron momentum and impact angle on the EMC, separately for the barrel and endcap parts of the EMC. 

The correction for PID is $C^\mathrm{PID} = 1.009\pm0.009\pm0.015,$ where the first error is due to the statistics of the CS and
the second is due to the incomplete kinematic coverage of the CS.

\subsection{Evaluation of trigger efficiency ratio}
The trigger efficiencies for \ke and \km are evaluated from data, by comparing two almost independent trigger algorithms based on DC and
EMC information. The correlation between these two triggers is topological in nature,
and has been evaluated from MC. The dependence of the ratio of trigger efficiencies on
the data taking period has been studied: the evaluated trigger efficiency ratio is stable well below the 1\% level, and is independent
of the run conditions. The ratio between the trigger efficiencies for the \ke
and \km processes is $\epsilon^\mathrm{TRG}_{K\mu2}/\epsilon^\mathrm{TRG}_{Ke2}=0.998\pm0.009\pm0.006,$ 
where again the first and second errors are statistical and systematic, respectively.

\subsection{Result}
Using the number of observed \ke and \km events and all
corrections as in Eq.~\ref{eq:ratio}, we get the preliminary result
\begin{equation}
R_K = (2.55\pm0.05\pm0.05)\times10^{-5}.
\end{equation}
This value is compatible within the error with the SM prediction, $R_K = (2.472\pm0.001)\times10^{-5}.$

\section{Prospects for improvement}
Three sources contribute to the present statistical uncertainty of 1.9\%: fluctuation in the signal counts (1.1\%), fluctuation in the
background to be subtracted (0.7\%), and statistical error on the MC estimate of the background (1.4\%). The dominant source is the latter, because the
selection efficiency for badly reconstructed \km events is lower in MC than in data, so that the MC background
under the \ke peak (the solid histogram of Fig.~\ref{ke2:fit}) had to be scaled by a factor of 4 to match the level of background in data.

Three improvements will be used to lower the statistical uncertainty on $R_K$: first, a factor of 30\% more data still have to be analyzed; 
second, MC production in progress will increase the MC statistics by a factor of two; third, various improvements will be performed to the selection
algorithm in order to increase the background rejection power, at least by a factor of two.

The uncertainty on the PID efficiency is 1.75\% and is the dominant contribution to the present systematic error.
The CS statistics will be improved by a factor of four and additional studies of PID methods are needed to reduce 
the purely systematic contribution from 1.5\% down to
less than 1\%.  The current uncertainty on the one-prong selection efficiency is 0.9\% and is dominated by the statistics of the control sample used
(only 10~pb$^{-1}$). Using additional statistics this error will be pushed down to below 0.5\%. 
Additional studies of the data/MC
agreement on trigger variables for reconstructed events are needed in order to increase the statistics of signal
events for trigger efficiency evaluation. 

Significant progress toward the 1\% goal is expected to come from the use of an additional sample of \ke events in which
the kaon decays before the inner DC wall. These events can be selected using tag information, by extrapolating the kaon trajectory known from the
tagging kaon from the IP to the point of closest approach with a secondary track reconstructed in the first layers of the DC.
This method has been used and provides an additional 37\% of \ke events. 
The systematic studies on this selection algorithm have yet to be completed.

\end{document}